\documentclass[a4paper,fleqn,usenatbib]{mnras}

\usepackage{newtxtext,newtxmath}

\usepackage[T1]{fontenc}
\usepackage{ae,aecompl}

\usepackage{graphicx}	
\usepackage{amsmath}	
\usepackage{amssymb}	

\newcommand{\cmc}{\,cm$^{-3}$}  

\title[$nl$-model for HRL masers]{An $nl$-model with radiative transfer for hydrogen recombination line masers}

\author[A. Prozesky and D. P. Smits]{
Andri Prozesky,$^{1}$\thanks{E-mail: prozea@unisa.ac.za (AP)}
and Derck P. Smits$^{1}$
\\
$^{1}$Department of Mathematical Sciences, University of South Africa, Private 
Bag X6, Florida 1709, South Africa\\
}

\date{Accepted XXX. Received YYY; in original form ZZZ}

\pubyear{2019}

\begin{document}
\label{firstpage}
\pagerange{\pageref{firstpage}--\pageref{lastpage}}
\maketitle

\begin{abstract}
Atomic hydrogen masers occur in recombination plasmas
in sufficiently dense \ion{H}{ii} regions. These hydrogen recombination line
(HRL) masers have been observed in a handful of objects to date and
the analysis of the atomic physics involved has been rudimentary.
In this work a new model of HRL masers is presented which uses an
$nl$-model to describe the atomic populations interacting with
free-free radiation from the plasma, and an escape probability framework
to deal with radiative transfer effects. The importance of including
the collisions between angular momentum quantum states and the free-free
emission in models of HRL masers is demonstrated. The model is used
to describe the general behaviour of radiative transfer of HRLs and
to investigate the conditions under which HRL masers form. The model
results show good agreement with observations collected over a broad
range of frequencies. Theoretical predictions are made regarding the
ratio of recombination lines from the same upper quantum level for these
objects.

\end{abstract}

\begin{keywords}
masers -- radiative transfer -- atomic processes -- line: formation --  ISM: atoms\\
\end{keywords}




\section{Background}
\subsection{Hydrogen recombination line masers}
\defcitealias{2018Prozesky}{PS18}

Astronomical masers occur when spectral lines are amplified through
stimulated emissions and are observed to be much brighter than expected
under local thermodynamic equilibrium (LTE) conditions. Molecular
astronomical masers have been a very useful tool to probe conditions
in a wide variety of sources. Traditional masers are produced by rotational
or vibrational transitions in various molecules, but recombination
line masers of hydrogen have been discovered in a handful of objects. 

\citet{1966goldberg} showed that recombination lines are amplified by stimulated emissions in the Rayleigh-Jeans limit, even at low optical depths. The theoretical possibility of hydrogen recombination line (HRL) masers was considered by \citet{1978Krolik}
to account for the anomalous hydrogen line intensities found in dense gasses associated with active galactic nuclei. The first cosmic high-gain HRL maser was discovered in the young stellar object MWC 349A \citep{1989Martin-Pintado,1989Martin-Pintado2}. This maser source has since been studied extensively  \citep{1992Planesas,1992Thum,1994Thum,1994Thum_b,1994Gordon,1994Martin-Pintado,1994PonomarevInterp,1998Thum,2001Gordon,2008Weintroub} and the evidence confirms the presence of strongly masing recombination lines. 

For some time, MWC 349A was the only source in which HRL masers have been
detected, but growing interest in the subject has prompted more searches, leading to 
the identification of a number of HRL masers in other objects. Masing HRLs have been 
observed in $\eta$ Carinae \citep{1995Cox}, in a high-velocity ionized jet from the 
star-forming region Cepheus A HW2 \citep{2011Jimenez-Serra}, from the ultra-compact 
\ion{H}{ii} region Mon R2 \citep{2013Jimenez-Serra}, in the planetary nebula
Mz 3 \citet{2018Aleman} and in the gas surrounding the B[e] star MWC 922 \citep{2017SanchezContreras}. Because of anomalous H30$\alpha$ line emission, \citet{2019Murchikova} suggest the presence of an HRL maser in the accretion disc around the central Galactic black hole Sgr A$^*$. The first extragalactic HRL maser has  been detected
in the star-forming galaxy NGC 253 \citep{2018Baez-Rubio}.  The prospects
of using HRL masers on cosmological scales to study the first galaxies
\citep{2013Rule} or the epoch of recombination and reionization \citep{1997Spaans} have been considered.

The environments in which these atomic masers can form are distinctly different
from those of their molecular counterparts. For recombination lines
to form, the emitting gas has to be ionized. For hydrogen this
generally requires a temperature of $\sim10^4$\,K. This is the characteristic temperature of a photoionized nebula, which will be the focus of this work. Collisionally ionized gasses can have significantly higher electron temperatures. The host clouds
of molecular masers are necessarily cooler than the molecule's dissociation
temperature and therefore are relatively cool. A population
inversion occurs in hydrogen over several decades of $n$-levels in a recombination nebula, whereas in molecular masers the inversion is often limited to a few levels. A result of this is that many adjacent HRL lines will exhibit masing behaviour at the same time instead of just a few specific lines as is the case with molecules. In HRL masers, the pumping scheme for the population inversions is a natural consequence of the capture-cascade processes in the atomic component of an ionized gas, which is discussed in more detail in \citet{1996Strelnitski}. In many molecular masers the details associated with the pumping scheme are unclear. 
It should also be noted that because hydrogen makes up the bulk of almost
all astronomical gasses, the masing species can be seen in very high
column densities in the case of hydrogen. For molecular masers the relevant constituents have low number densities compared to the H$_2$ content.

\subsection{Challenges of maser modeling}

Modeling the interaction between line photons and the emitting matter
in an astronomical cloud is key to understanding masers. Such a model
will have to account for not only the recombination theory and line
formation in multi-level atoms, but also deal with the radiative
transfer of the line photons through the cloud. A complete and simultaneous
description of both of these components is a complex problem.

A capture-collision-cascade (C$^{3}$) model, such as the one described
by \citet{2018Prozesky} (hereafter \citetalias{2018Prozesky}), accounts for the influence that radiation
has on atoms in an astronomical cloud by solving the statistical balance
equations (SBE) for the level populations of hydrogen.
This type of model is appropriate when the plasma is optically
thin. When maser action is present, the line radiation is optically thick and radiative transfer effects have to be accounted for. The equation of radiative
transfer (ERT) contains the effects that the matter has on the radiation
field as it travels through the medium. However, the coefficients
in the ERT depend on the local level populations, and the level populations
in turn are influenced by the intensity of the radiation field, which
is a non-local quantity. This means that these two effects are coupled
to one another and, in principle, have to be solved simultaneously in
a self-consistent manner to obtain the theoretical intensities of
the spectral lines escaping the cloud. Simplifying assumptions, such as
the escape probability approximation (EPA) discussed in section \ref{sec:The EPA}, are often employed to make the problem tractable.

Hydrogen masers are simpler than molecular masers in the extent that
the calculations include all relevant atomic processes to create population inversions. However, hydrogen masers are more complex to model in the sense that several atomic levels that interact directly with one another exhibit
masing at the same time. The maser action in hydrogen therefore has
a complex effect on the level populations of the participating levels.

\subsection{Previous models of HRL masers}

There have been some endeavours to construct
a theoretical framework for HRL masers. \citet{1990Walmsley} extended
the departure coefficient calculations of \citet{1977Brocklehurst}
to higher densities in response to the discovery of the first HRL
masing region. \citet{1996Strelnitski} addressed the theoretical
foundations of HRL masers and considered conditions necessary for their
formation.

Most theoretical models for HRL masers have focused on the morphology
of the emitting region \citep[e.g.\!][]{1994PonomarevModel,1996StrelnitskiB,2008Weintroub}
and it has been suggested that the masing is strongly related to the
structure and kinematics of the emitting gas \citep{2002Martin-Pintado}.
Most notable is the three-dimensional non-LTE radiative transfer code
MORELI \citep{2013Baez-Rubio}. MORELI uses pre-calculated departure
coefficients of either \citet{1990Walmsley} or \citet{1995Storey},
but does not solve the SBE in a self-consistent way.

\citet{2000Hengel} incorporated radiative transfer effects into a
C$^{3}$ model to assess the effects of saturation on the level populations.
They employed an $n$-model which neglects the effects of the elastic collisions
between angular momentum states, as opposed to an $nl$-model
in which they are included. \citet{2000Hengel} found that the effects
of the radiative transfer on the level populations of hydrogen are
important. 

\subsection{Outline of this paper}

Section~\ref{sec:Radiative-transfer} gives a quick overview of the
basic equations involved in radiative transfer theory and establishes
the notation used throughout this paper. The importance of including
the effects of both the free-free radiation field and elastic collisions
for HRL maser models are highlighted in section ~\ref{sec:Physical-considerations}.
A summary of the EPA that was used in the current
model is given in section~\ref{sec:The EPA}, as well as a discussion
regarding its strengths and limitations. Section~\ref{sec:The-model} outlines the calculational details of
the HRL maser model used here. The main results, as well as comparison
with observations and predictions are given in section~\ref{sec:Results}.
The main conclusions are summarized in section~\ref{sec:Conclusions}.


\section{Radiative transfer}
\label{sec:Radiative-transfer}

The ERT describes the radiation added to and subtracted from a given
ray as it travels through a medium and is given by 
\begin{equation}
\frac{\mathrm{d}I_{\nu}}{\mathrm{d}l}=-\kappa_\nu I_\nu+j_\nu,\label{eq:ERT}
\end{equation}
where $I_{\nu}$ is the specific intensity and $l$ is the path along the ray. The net (line + continuum) volume emission and absorption
coefficients at the frequency $\nu$ are given by $j_{\nu}$ and $\kappa_{\nu}$, respectively. The source function is defined as $S_{\nu}=j_{\nu}/\kappa_{\nu}$.

The total line emission coefficient $j_{nm}$ describes radiation
added to the radiation of the spectral line of the transition $n\rightarrow m$
through spontaneous emissions and is defined as 
\begin{align}
j_{nm} & =\frac{h\nu}{4\pi}\sum_{l=0}^{n-1}\sum_{l^{'}=l\pm1}b_{nl}N_{nl}^{*}A_{nl,ml^{'}}\label{eq:line em coef}
\end{align}
where $h$ is Planck's constant, $b_{nl}$ is the departure coefficient of level $nl$, $N_{nl}^{*}$
is the population of level $nl$ in LTE and $A_{nl,ml^{'}}$
is the Einstein A-value for the $nl\rightarrow ml^{'}$ transition.

The total line absorption coefficient $\kappa_{mn}$ gives the contribution
of stimulated emissions ($B_{nm}$) and absorptions ($B_{mn}$) to
the emerging radiation field as 
\begin{align}
\kappa_{mn} & =\frac{h\nu}{4\pi}\left(N_{m}B_{mn}-N_{n}B_{nm}\right)\label{eq:line abs coef def}\\
 & =\frac{h\nu}{4\pi}\sum_{l=0}^{n-1}\sum_{l^{'}=l\pm1} b_{ml^{'}}N^{*}_{ml^{'}}B_{ml^{'},nl}\left(1-\frac{b_{nl}}{b_{ml^{'}}}\mathrm{e}^{-h\nu/kT_\mathrm{e}}\right)\,,\label{eq:line abs coef}
\end{align}
where $k$ is the Boltzmann constant and $T_\mathrm{e}$ is the temperature of the free electron gas, which is assumed to have a Maxwellian distribution.

From the definition in equation~(\ref{eq:line abs coef def}), it is clear
that $\kappa_{mn}$ can become negative if the number of stimulated
emissions exceeds the number of absorptions, thereby increasing the line intensity. The term inside brackets in equation~(\ref{eq:line abs coef}) is the correction for stimulated emission.

Each transition has a well defined frequency $\nu_{nm}$ associated
with it, but in practice there is a range of frequencies around $\nu_{nm}$
where photons from the line transition can be either emitted or absorbed.
The emission and absorption coefficients at any particular frequency
within the line are described by 
\begin{equation}
j_{nm}^{\nu}=j_{nm}\phi_{\nu}\quad\mathrm{and}\quad\kappa_{mn}^{\nu}=\kappa_{mn}\phi_{\nu}\,.\label{eq:integrated abs emm}
\end{equation}
Strictly speaking, the line profile functions $\phi_{\nu}$ for emissions
and absorptions are not equal, but they are similar enough for the
purposes of this discussion that they will be considered to be equal.

At low enough frequencies, when the continuum is significant which usually occurs in the Rayleigh-Jeans limit, the line and continuum radiation are formed together. This means
the net quantities (indicated by subscripts $\nu$) in equation~(\ref{eq:ERT}) must
take into account the contributions of both the line radiation and
the continuum (indicated by subscripts $\mathrm{c}$), so that 
\begin{equation}
\kappa_{\nu}  =\kappa_{mn}^{\nu}+\kappa_{\mathrm{c}},\qquad j_{\nu}=j_{nm}^{\nu}+j_{\mathrm{c}}\,.
\end{equation}
The net source function $S_{\nu}$ is given by
\begin{equation}
S_{\nu}=\frac{j_{nm}^{\nu}+j_{\mathrm{c}}}{\kappa_{mn}^{\nu}+\kappa_{\mathrm{c}}}\,.\label{eq:source function}
\end{equation}
For a homogeneous medium of thickness $L$ the optical depth is given
by
\begin{equation}
\tau_{\nu}  =  -L\kappa_{\nu}\,.\label{eq:opkappa}
\end{equation}

It is usual to calculate the line intensity at a specific frequency by subtracting the continuum from the total intensity using 
\begin{equation}
\bar{J}^\nu_{nm} = S_{\nu}\left(1-e^{-\tau_{\nu}}\right)-B_{\nu}\left(1-e^{-\tau_{\mathrm{c}}}\right)\, ,\label{eq:opthin_Inm}
\end{equation}
where $B_\nu$ is the Planck distribution function, see for example \citet{1966goldberg} and \citet{1996Strelnitski}. However, if the gas becomes  optically thick, the continuum photons will interact with the atoms in a significant way so that the contribution of the line photons to the total intensity at line centre is
\begin{equation}
J_{nm}^\nu=\left( \frac{j_{nm}^{\nu}}{\kappa_{mn}^{\nu}+\kappa_\mathrm{c}} \right)  \left( 1- e^{-\tau_{\nu}}\right) \,.\label{eq:opthick_Inm}
\end{equation}
Similarly, the contribution of the continuum photons will be
\begin{equation}
J_\mathrm{c}^\nu=\left(  \frac{j_\mathrm{c}}{\kappa_{mn}^{\nu}+\kappa_\mathrm{c}} \right)  \left( 1- e^{-\tau_{\nu}}\right) \,.\label{eq:opthick_Ic}
\end{equation}
Equations~(\ref{eq:opthick_Inm}) and (\ref{eq:opthick_Ic}) account for the fact that the mean free path of a photon depends only on its frequency, not on its origin. The line intensity $J_{nm}^\nu$ described by equation~(\ref{eq:opthick_Inm}) cannot be measured directly from a spectrum. Therefore the quantity $\bar{J}^\nu_{nm}$ described by equation~(\ref{eq:opthin_Inm}) will be referred to as the observable line intensity under optically thick conditions.

From an observational perspective, the quantity described by equation~(\ref{eq:opthin_Inm}) can be extracted from an observed spectrum, even if the radiation is emitted from an optically thick region where maser effects are important. A comparison between the behaviour of $\bar{J}^\nu_{nm}$ and $J_{nm}^\nu$ is discussed in section \ref{subsec:gen_trends}. 

In our models, the continuum absorption coefficient $\kappa_{\mathrm{c}}$ in the
Rayleigh-Jeans regime is calculated  using the expression of \citet{1961Oster} given by
\begin{equation}
\kappa_{\mathrm{c}}=\left(\frac{N_\mathrm{e}N_{\mathrm{i}}}{\nu^{2}}\right)\left(\frac{8Z^{2}e^6}{3\sqrt{3}m_\mathrm{e}^{3}c}\right)\left(\frac{\pi}{2}\right)^{1/2}\left(\frac{m_\mathrm{e}}{k T_\mathrm{e}}\right)^{3/2}\left\langle g\right\rangle \,,
\label{eq:kappa_c}
\end{equation}
where $N_\mathrm{e}$ and $N_\mathrm{i}$ are the number densities
of the electrons and ions, $Z$ is the atomic charge, $e$ is the
elementary charge, $m_\mathrm{e}$ is the electron mass, and $c$ is
the speed of light. For $T_\mathrm{e} < 550\,000$\,K,
the Gaunt factor averaged over a Maxwellian velocity distribution
\citep{1961Oster} can be approximated by 
\begin{equation}
\left\langle g\right\rangle \approx\frac{\sqrt{3}}{\pi}\ln\left[\left(\frac{2k T_\mathrm{e}}{\gamma m_\mathrm{e}}\right)^{3/2}\frac{m_\mathrm{e}}{\pi\gamma Ze^{2}\nu}\right]\;,
\label{eq:gaunt}
\end{equation}
where $\gamma$ is the exponential of the Euler--Mascheroni constant.
The continuum emission coefficient $j_\mathrm{c}=\kappa_\mathrm{c}B_\nu(T_\mathrm{e})$.


\section{Physical considerations}
\label{sec:Physical-considerations}

\subsection{Angular momentum changing collisions}

\citet{1996Strelnitski} present some of the theoretical aspects of
high gain HRL masers. They specifically focus on the optimum electron
density for each maser line, i.e.\ the density at which the magnitude
of the negative absorption coefficient $\left|\kappa_{\nu}\right|$
is maximized. In these calculations, they use departure coefficients
derived from an $n$-model. They argue that the densities where
maser lines are formed are high enough ($>10^7$\cmc)
that the angular momentum structure of the atoms can be ignored because the
inelastic collisions will set up Boltzmann distributions among the $l$-levels. This sentiment has been echoed by \citet{2000Hengel}.

However, at the levels with $n\lesssim 30$ for $T_\mathrm{e}=10^4$\,K
and $N_\mathrm{e}=10^8$\cmc\ radiative recombination is the dominant process governing the level populations of the low $l$-states. For most atomic levels, radiative recombination will be orders of magnitude faster than the next fastest process, three-body recombination. Radiative recombination disrupts the Boltzmann distributions, since it highly favours low $l$-levels over high values of $l$ and there will not be Boltzmann distributions in the angular momentum states. The effect of accounting for the $l$-structure of the atoms is explored in section \ref{subsec:optinvs}.

\subsection{Free-free emission}

 In the case of a homogeneous pure hydrogen gas, the intensity of the free-free emission generated by the electrons
within a plasma is proportional to  $N_\mathrm{e}^2$. 
\citetalias{2018Prozesky} showed that the effects of the free-free emission
on level populations become increasingly important as the electron
density increases. Therefore, it is important to include the free-free emission
in model calculations of HRL masers that necessarily occur at high
densities.

\citet{1996Strelnitski} used departure coefficients from an $n$-model without free-free
radiation included to draw their conclusions regarding conditions
in which hydrogen masers will form. The masers with the highest gain are calculated by considering the magnitude of the net absorption coefficients
under various conditions. Their results have been found to be consistent
with observations \citep{1998Thum,1996StrelnitskiB}

In contrast, the preceding discussion indicates that the angular momentum
changing collisions and the free-free emissions should not be neglected
at these densities. Both effects change the energy levels for which masing
is possible by 5 to 10 levels, but at $N_\mathrm{e}>10^6$\cmc\ 
the two effects counteract one another when only considering $\left|\kappa_{\nu}\right|$.
The inclusion of the angular momentum changing collisions narrows
the range of possible $n$-levels from which H$n\alpha$ masers are
possible whereas including the free-free emission widens the range. Neither
effect changes the maximum value of $\left|\kappa_{\nu}\right|$ significantly,
only the values of $n$ where $\kappa_{\nu}<0$ are altered. The partial
cancellation of the two effects leads to the results
of \citet{1996Strelnitski} being consistent with a more sophisticated
analysis.


\section{The escape probability approach}
\label{sec:The EPA}

\subsection{Basic theory}
\label{subsec:Basic-theory}

A popular simplifying assumption when doing recombination line calculations
is to assume that the emitting cloud is either completely opaque or
completely transparent to particular lines. For example, the widely
used Case B of \citet{1938BakerMenzel} assumes that all Lyman transitions
are optically thick, whereas all others are optically thin. This assumption
is very easy to incorporate into calculations and has been found
to work well for nebular conditions where densities are low \citep{1962Osterbrock}.

When solving a C$^{3}$-type model, such as described in \citetalias{2018Prozesky},
it is standard to use the Case A/B assumption. In the case where line
radiation is assumed to be optically thin ($\tau\ll1$), diffuse radiation
is assumed to escape the cloud without interacting with the particles.
In the other extreme where the cloud is completely optically thick
to all line radiation ($\tau_{\nu}\gg1$), all the level populations
will follow Boltzmann distributions and the mean intensity $J_{\nu}=S_{\nu}$.

The EPA addresses the situation between these two extremes, where
a portion of the radiation is trapped in the cloud and some of it
is allowed to escape. If the fraction of photons with frequency $\nu$
that escape the cloud is labeled $\beta_{\nu}$, then $\left(1-\beta_{\nu}\right)$
of the photons will be reabsorbed by the medium. With the fraction
$\left(1-\beta_{\nu}\right)$ of emitted photons trapped in the medium,
the mean intensity can be approximated by (ignoring the continuum
radiation fields for the moment) 
\begin{equation}
J_{\nu}=\left(1-\beta_{\nu}\right)S_{\nu}\,.\label{eq:J_EPA}
\end{equation}

For maser transitions which have inverted level populations $\beta_{\nu}>1$. Strictly speaking, $\beta_{\nu}$ depends on the full solution
of the ERT and cannot be calculated locally. However, if an approximation
can be derived that depends only on the geometry and local properties
of the cloud and is independent of intensity, then the original problem
is greatly simplified.

The EPA has been used extensively to model molecular masers. A popular
form of the escape probability is the large velocity gradient approximation
(see for example \citet{1997Sobolev,2002Cragg,1984Langer,2001Humphreys}).
For a spherically symmetric, homogeneous cloud in which the expansion velocity
is proportional to the radius, the escape probability becomes
\begin{equation}
\beta_{\nu}=\frac{1-\exp\left(-\tau_{\nu}\right)}{\tau_{\nu}}\,.\label{eq:beta_LVG}
\end{equation}

Another feasible escape probability for masers is 
\begin{equation}
\beta_{\nu}=\mathrm{e}^{-\tau_{\nu}}\label{eq:beta_HK}
\end{equation}
as used for example by \citet{1979Kegel,1980Koeppen,1984Chandra,1999Rollig}.
The authors argue that this form of $\beta_{\nu}$ does not make
any additional assumptions regarding the geometry or the changes in
transfer effects throughout the line profile and is therefore more
appropriate to use if these details are not known. In this work equation~(\ref{eq:beta_HK}) is used, as the aim is to derive general trends
and this form is more widely applicable. This results in the total mean intensity calculated by the model reducing to the sum of  equations~(\ref{eq:opthick_Inm}) and (\ref{eq:opthick_Ic}). It should be noted that the
results derived from using equation~(\ref{eq:beta_LVG}) are very similar
to that of equation~(\ref{eq:beta_HK}).

\subsection{Strengths and limitations}

EPA methods are frequently applied to astronomical problems when
the interest lies in overall line intensities emitted from a cloud
and not the exact details of line formation within the cloud. It is
computationally easy to implement and produces results which are in
good agreement with much more sophisticated and calculationally complex
methods \citep{1992Elitzur}. Notable radiative transfer codes such
as Cloudy \citep{2013Ferland}, XSTAR \citep{2001Kallman} and RADEX
\citep{2007vandertak} all make use of the EPA in various ways. However,
the method is approximate and its shortcomings should be taken into
consideration when it is used.

EPA methods give an overall approximation of the global properties
of a source. The level populations calculated in this formalism
are independent of location in the source and yield the mean level populations
that are consistent with the overall emission. Importantly, the resulting
level populations are consistent with saturation effects. Saturation occurs when the population difference between two levels of a maser transition is appreciably affected by the maser radiation.  Therefore,
the overall manner in which the maser emission affects the level populations is
accounted for.

Uniform physical conditions throughout the source are a built-in assumption
in this scheme, specifically a constant source function. However, the mean
intensity is position-dependent through the optical depth. Because the source function also depends indirectly
on $J_{\nu}$, this is
inherently contradictory. \citet{1992Elitzur} argues that masers are particularly
suitable to be treated in the EPA, due to the fact that maser source functions are essentially constant.

\citet{1990Elitzur} emphasizes the importance of the effects of beaming
when modeling masers. However, this is not a simple issue as discussed
in \citet{1992Lockett} and the EPA can only account for this in a
very approximate way. Because no particular geometry is assumed, beaming effects are neglected in this work.

Another limitation of the EPA is that it does not produce details about
line shapes, only line integrated quantities. The line profile
of maser emission will change depending on the level of saturation
\citep{1994Elitzur}. The details of this are lost in the EPA approach.

More accurate models for radiative transfer for multi-level atoms
do exist. The gold standard is the accelerated $\Lambda$-iteration
(ALI) method \citep{1991Rybicki} that does a much more detailed treatment
of radiative transfer, but is also very computationally expensive.
More recently, the coupled escape probability (CEP) method has been developed
\citep{2006Elitzur,2018AsensioRamos}. The CEP method rivals the ALI
methods in accuracy, but is much simpler to implement and faster to
compute. However, neither method accounts for saturation or beaming
effects correctly and the CEP method defers to the more rudimentary EPA if saturation effects
are important \citep{2018AsensioRamos}. \citet{2018Gray} and \citet{2019Gray} present a three-dimensional model based on CEP that accurately accounts for beaming and saturation effects, but currently only under a two-level approximation.

All EPA formalisms are derived from plausibility arguments and not
from first principles. This means that they provide no internal error
estimate and their accuracy can only be determined when their results
are compared to those of full radiative transfer treatments like the
ALI. \citet{2003Dumont} compared EPA and ALI results with specific
focus on AGN and X-ray binaries and found
that the EPA overestimates line intensities.
\citet{2016Nesterenok} also compared results from one-dimensional
EPA and ALI models for methanol masers and found that the EPA is
accurate if the cloud dimensions are large enough. Neither of these results are
directly transferable to the EPA model presented here for hydrogen,
but the limitations of this method should be kept in mind.


\section{The model}
\label{sec:The-model}

\subsection{Overview}

The EPA model used here is similar to that of \citet{2000Hengel} with
some important improvements. Most importantly, it includes the effects
of the elastic collisions so the calculations are done with a full
$nl$-model. Also, the effects of free-free radiation on the level
populations have been incorporated. Nevertheless, we use the same
form of the escape probability and general calculational approach.

Our atomic model is based on the C$^3$ $nl$-model described in \citetalias{2018Prozesky}
that was adapted to incorporate radiative transfer using the EPA as
described in section \ref{subsec:Basic-theory}. All atomic rates
are as described in \citetalias{2018Prozesky}. The iterative solver, as
opposed to the direct solver, was used to obtain the $b_{nl}$ values.
This streamlined the calculations which  had to be repeated many
times for increasing path lengths. The value of $n$ up to where the
$nl$-model was calculated was increased considerably from what was
used in \citetalias{2018Prozesky}, because the $n$-model results became
unreliable for large path lengths. This is discussed further in section
\ref{subsec:General-results}.

A box profile with the same amplitude as the Doppler profile is assumed for all lines so  that 
\begin{equation}
\phi_\nu=\begin{cases}
\frac{1}{\sqrt{\pi}\nu_{D}} & \mathrm{for\:}\left|\nu-\nu_{0}\right|<\frac{\sqrt{\pi}\nu_{D}}{2}\\
0 & \mathrm{otherwise}
\end{cases}\label{eq:box profile}
\end{equation}
where 
\begin{equation}
\nu_{D}=\frac{\nu_0}{c}\sqrt{\frac{2kT_\mathrm{e}}{m_\mathrm{e}}}\,.
\end{equation}

The SBE as described in \citetalias{2018Prozesky} are not a closed system. Details
regarding the ionizing radiation are not specified and therefore the population of the ground state cannot be calculated. In addition to the
SBE, the degree of ionization $\zeta = N_0/N_\mathrm{e}$, where $N_0$
is the number density of neutral hydrogen, is fixed to $\zeta=10^{-4}$.
The total number density of neutral hydrogen is equal to the number
densities of atoms with electrons in all bound states, so that 
\begin{equation}
b_1 = \frac{N_\mathrm{e}}{N_1^*}\left( \zeta - \sum_{n=2} \frac{b_n N_n^*}{N_\mathrm{e}}\right)\,.
\end{equation}

\subsection{Calculational procedure}

First, the SBE are solved in the optically thin case ($\beta_{\nu}=1$
for all lines) to yield departure coefficients that are equivalent
to the ones given in \citetalias{2018Prozesky} for Case A. From the calculated values of $b_{nl}$, the net
absorption coefficients, emission coefficients and source functions
are calculated for each line using equations~(\ref{eq:line em coef}),
(\ref{eq:line abs coef}) and (\ref{eq:source function}).

The path length $L$ is then increased and the optical depths and
escape probabilities are calculated from equations~(\ref{eq:opkappa})
and (\ref{eq:beta_HK}), respectively. Since the escape probabilities
become unstable near $\tau_{\nu}=0$, a Taylor expansion is used for
small values of $\tau_{\nu}$. The resulting mean intensities are
calculated for each line using 
\begin{equation}
J_{\nu}=\left(1-\beta_{\nu}\right)S_{\nu}\,.
\end{equation}
 The SBE are solved
again with these values of $J_{\nu}$ incorporated into the rates
of the stimulated processes. The process is repeated for increasing
values of $L$.

Because the atomic rates on which the departure coefficients depend have inherent inaccuracies, the $b_{nl}$ values cannot be calculated to an arbitrary precision.
The line absorption coefficients depend on the ratios of departure
coefficients and therefore are highly sensitive to numerical errors
in the $b_{nl}$'s. At high column densities, the line absorption
coefficients become unstable and exhibit oscillating behaviour. This
occurs abruptly as iterations over $L$ are performed.

Small inaccuracies in the $b_{nl}$s are amplified by the strong
dependence of both $\beta_{\nu}$ and $J_{\nu}$ on $\kappa_{mn}$
as the iterative process progresses. This is especially true in the
region where $\left|\kappa_{mn}\right|>\kappa_\mathrm{c}$ where
masing can occur. The procedure is stable up until the point where
the $\kappa_{n\,,n+1}$ start to show minor oscillating behaviour,
which is exaggerated to unphysical results within a few iterations.
The point in the procedure where the results become unstable has
been extended by applying a moving average filter to both the $\kappa_{n+1,\,n}$
and $J_{\nu}$ for H$n\alpha$ transitions at large values of $L$. This is a common method
used for smoothing data affected by noise. In this case, the mean of blocks
of 3 points are calculated in the region where $\kappa_{n\,,n+1}<0$.
For a data set $x_{i}$, the smoothed data is given by
\begin{equation}
\bar{x}_{i}=\frac{1}{3}\sum_{k=i-1}^{i+1}x_{k}\,.
\end{equation}

The breakdown of the model at high column densities is not that limiting
in practice when compared to available observations. The intensities
of H$n\alpha$ transitions from MWC 349A are well within the limits
of this model (see section \ref{subsec:comp-obs}).


\section{Results and discussion}
\label{sec:Results}

\subsection{Conditions for hydrogen masers}
\label{subsec:General-results}

The pumping mechanism for hydrogen masers is the ionization-recombination
cycle, so it is important that the gas is ionized. Canonically, ionized
nebulae are taken at $T_\mathrm{e}\sim10^4$, but mechanisms such
as bright forbidden line emission due to high metallicity can cool the
gas while keeping the hydrogen mostly ionized. If the electron temperature 
is too high, the interactions
between the free and bound electrons become very fast and the populations
of the bound electrons thermalize. A temperature range of 
$3\,000\,\mathrm{K}\leq T_\mathrm{e}\leq 15\,000$\,K
was considered here.

Spectral lines will exhibit high-gain maser action if the conditions
are such that stimulated emissions become the dominant atomic process
and $\tau_{\nu}<-1$. This requires a large column density along the
line of sight, which can be achieved either with high number densities
of hydrogen atoms or long path lengths. The model results show that a path length
of $L\sim10^{16}$\,cm is required to produce maser action at
a density of $10^6$\cmc. Each order of magnitude decrease in $N_\mathrm{e}$
results in about an order of magnitude increase in $L$ to produce a maser.
Maser action requires velocity coherence along the amplification path, which 
puts an upper limit on $L$. Therefore, electron densities $N_\mathrm{e} \geq 10^7$\cmc\
will be considered here.

For densities higher than $N_\mathrm{e} \sim 10^{10}$\cmc\ the net absorption 
coefficient is negative for $n \leq 10$, so that masing is theoretically possible 
for these lines. However, the population of the ground level is set artificially 
to a fixed level of ionization in this model, so the results for the lowest 
$n$-levels are probably not very accurate and densities above this threshold are 
not considered. It is also unlikely that a gas of this density will be sufficiently
ionized for recombination lines to appear.

 \subsection{General trends}
 \label{subsec:gen_trends}

\begin{figure}
	\includegraphics[width=\columnwidth]{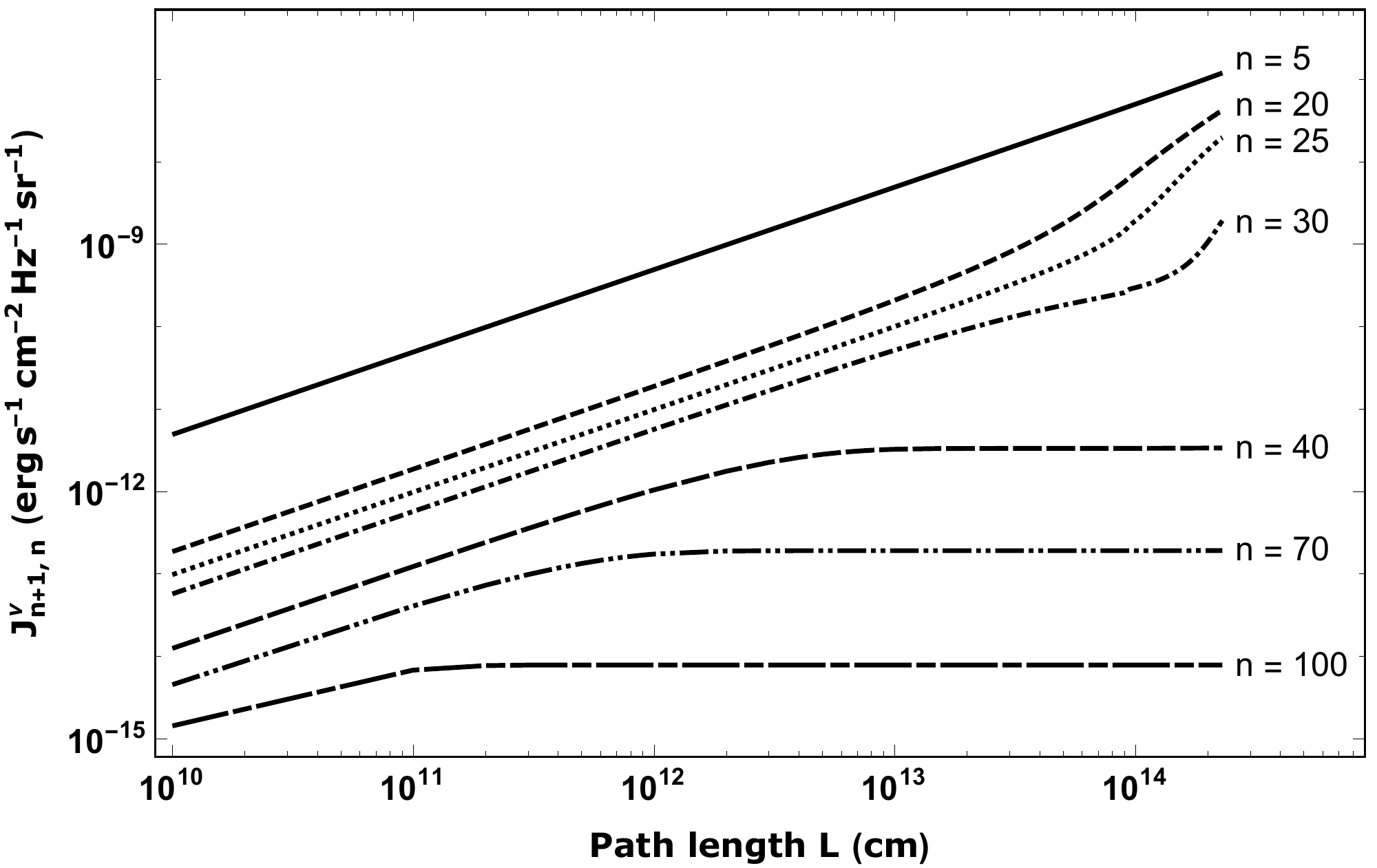}
		\caption{The change of the intensities at line centre for
various H$n\alpha$ transitions with path length for a gas at $T_\mathrm{e}=10^4$\,K
and $N_\mathrm{e}=10^8$\cmc.}
\label{fig:n_evol}
\end{figure}

Fig. \ref{fig:n_evol} shows how the  line centre intensities, calculated using the first term of equation~(\ref{eq:opthick_Inm}), change
as the path length is increased in the model for various H$n\alpha$
transitions. For small path lengths, the intensities of all lines increase
linearly with path length, as is expected for optically thin lines.
The behaviour of the line intensities change at the point when $\left|\tau_{\nu}\right|>1$ in
one of two ways, depending on whether $\tau_{\nu}$ is positive or
negative.

For the conditions shown in Fig. \ref{fig:n_evol}, the lines with
$n\geq40$ have positive optical depths which increase as $n$ increases 
throughout the iterations over path length. The level populations for
these lines are not strictly inverted, but are ``overheated'' as
discussed in \citet{1996Strelnitski}. The upper level of these lines
are overpopulated with respect to the LTE populations so that the lines
are still enhanced by stimulated emissions even though the absorption
coefficients for these lines are positive. These line intensities
increase linearly with path length until the cloud becomes larger
than their characteristic path length (for which $\tau_{\nu}=1$).
Once they are optically thick, their intensities remain constant as
the size of the cloud is increased, because the lines cannot be observed
from deeper in the cloud than their characteristic path length. Because 
$\tau_{\nu}$ for the H$n\alpha$ lines increases with $n$, the lower frequency 
(higher $n$) lines become optically thick before the lower $n$ lines as 
$L$ increases.

The optical depths of the H$20\alpha$ and H$25\alpha$ lines are
negative and their intensities start to increase exponentially with
distance once $\tau_{\nu}<-1$ for their respective optical depths
and maser action sets in. The case of H$30\alpha$ is interesting, because 
its optical depth is positive for small path lengths, but it is ``attracted'' 
into the masing range as masing in adjacent lines become effective. This 
phenomenon is discussed in more detail in section \ref{subsec:optinvs}. We 
do not expect the exponential growth to continue indefinitely, 
but because the model becomes unstable as the path length increases we cannot 
investigate this region of phase space. The optical depth of
the H$5\alpha$ line is also negative, but $\tau_{\nu}>-1$ for path 
lengths $L < 10^{15}$\,cm which is where this model is terminated.

\begin{figure}
	\includegraphics[width=\columnwidth]{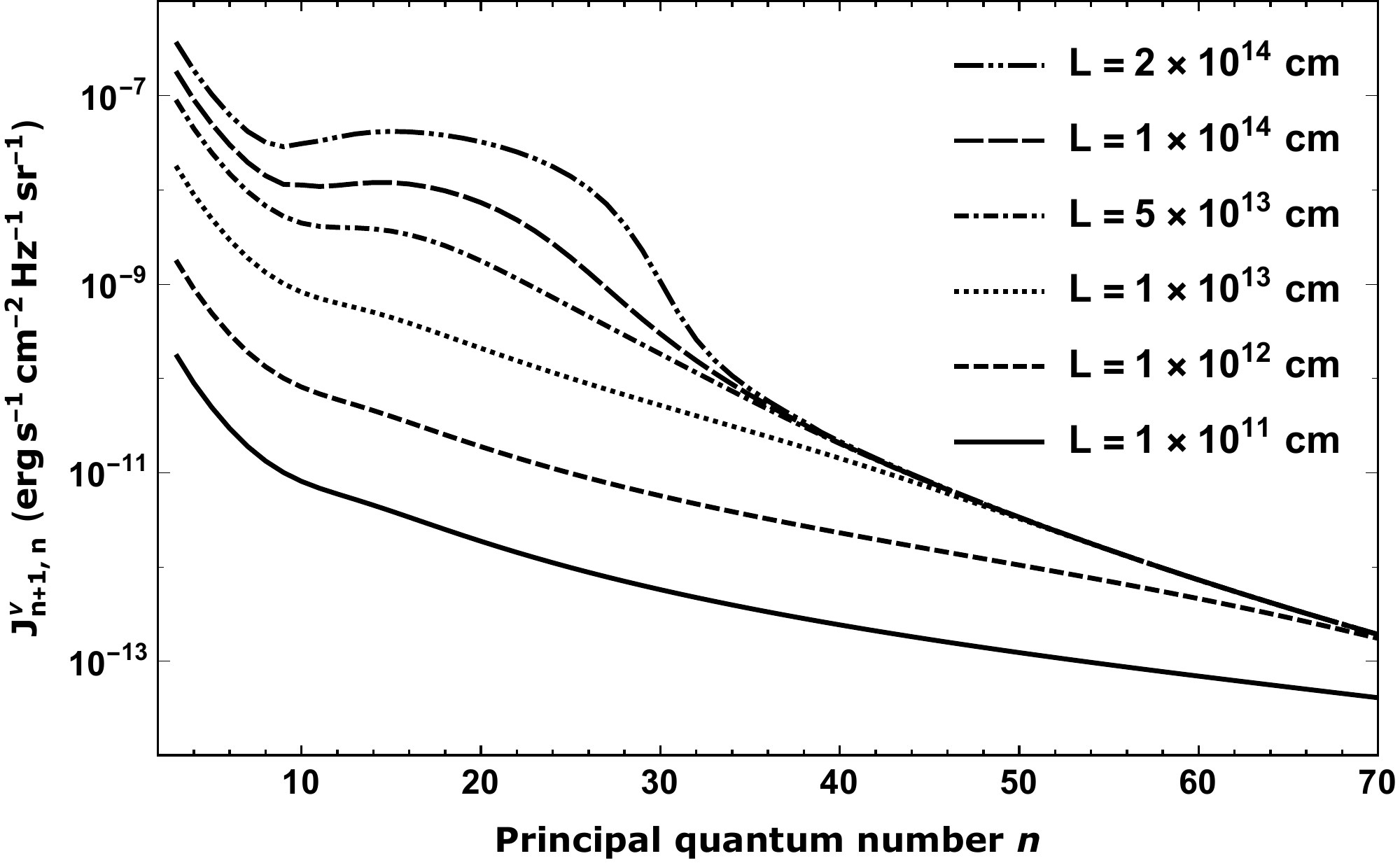}
		\caption{Intensities at line centre of H$n\alpha$ transitions
with respect to principal quantum number for a gas with $T_\mathrm{e}=10^4$\,K
and $N_\mathrm{e}=10^8$\cmc\ for different path lengths.}
\label{fig:bump_dist}
\end{figure}

Fig.~\ref{fig:bump_dist} shows the line intensities for
H$n\alpha$ lines in a gas with $T_\mathrm{e}=10^4$\,K
and $N_\mathrm{e}=10^8$\cmc\ as a function of $n$
for different path lengths. The intensities increase for all $n$ from
the optically thin values as $L$ is increased, as is also illustrated
in Fig. \ref{fig:n_evol}. As $L$ increases, the lines get optically
thick from high values of $n$ and then do not increase further.
If the path length becomes large enough that $\tau_{\nu}<-1$ for
some lines, a bump starts to appear, indicating maser action in those 
lines. As the path length is increased further, the intensities
of the masing lines increase significantly with $L$, making the bump
more pronounced.

\begin{figure}
	\includegraphics[width=\columnwidth]{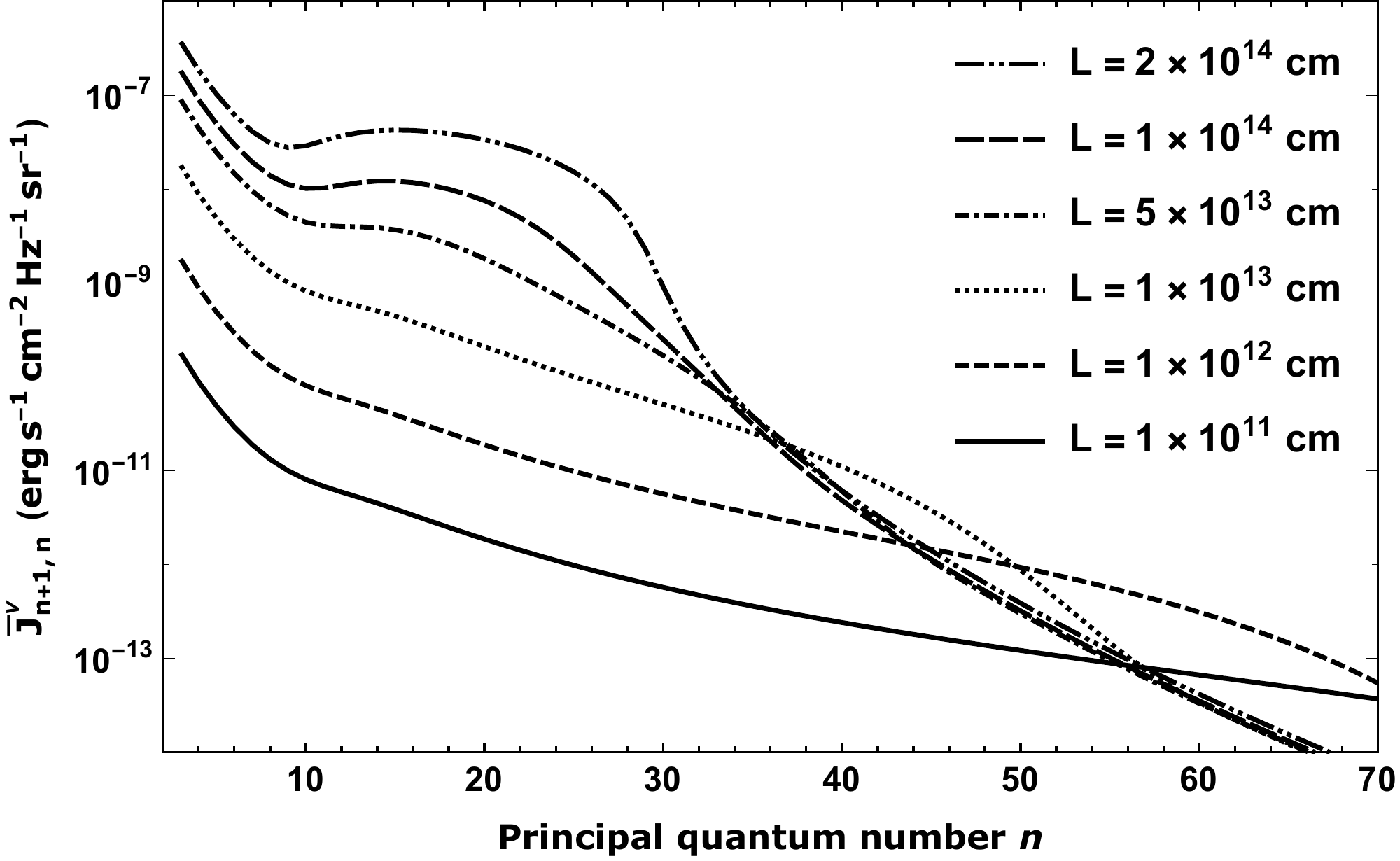}
		\caption{Observable intensities at line centre of H$n\alpha$ transitions $\bar{J}^\nu_{n+1,n}$
with respect to principal quantum number for a gas with $T_\mathrm{e}=10^4$\,K
and $N_\mathrm{e}=10^8$\cmc\ for different path lengths.}
\label{fig:bump_dist_obs}
\end{figure}

The observable intensities, as calculated using equation~(\ref{eq:opthin_Inm}) are 
shown in Fig.~\ref{fig:bump_dist_obs} for comparison with Fig.~\ref{fig:bump_dist}. The quantity $\bar{J}^\nu_{n+1,n}$ does not behave as one would expect for the intensity of a spectral line. For example, the brightness of $\bar{J}^\nu_{61,60}$ is \textit{decreased} if the path length along the line of sight is increased from $L=10^{12}$ cm to $L=10^{14}$ cm.

The discrepancy between $\bar{J}^\nu_{n,m}$ and $J^\nu_{n,m}$ is due to the  photons within the width of the line all experiencing an optical depth $\tau_\nu$, regardless of whether they were emitted by atomic transitions or as part of the continuum by the free electrons. Equation~\ref{eq:opthin_Inm} assumes that continuum photons are isolated from the atoms in the gas and do not interact with them.

\begin{figure}
	\includegraphics[width=\columnwidth]{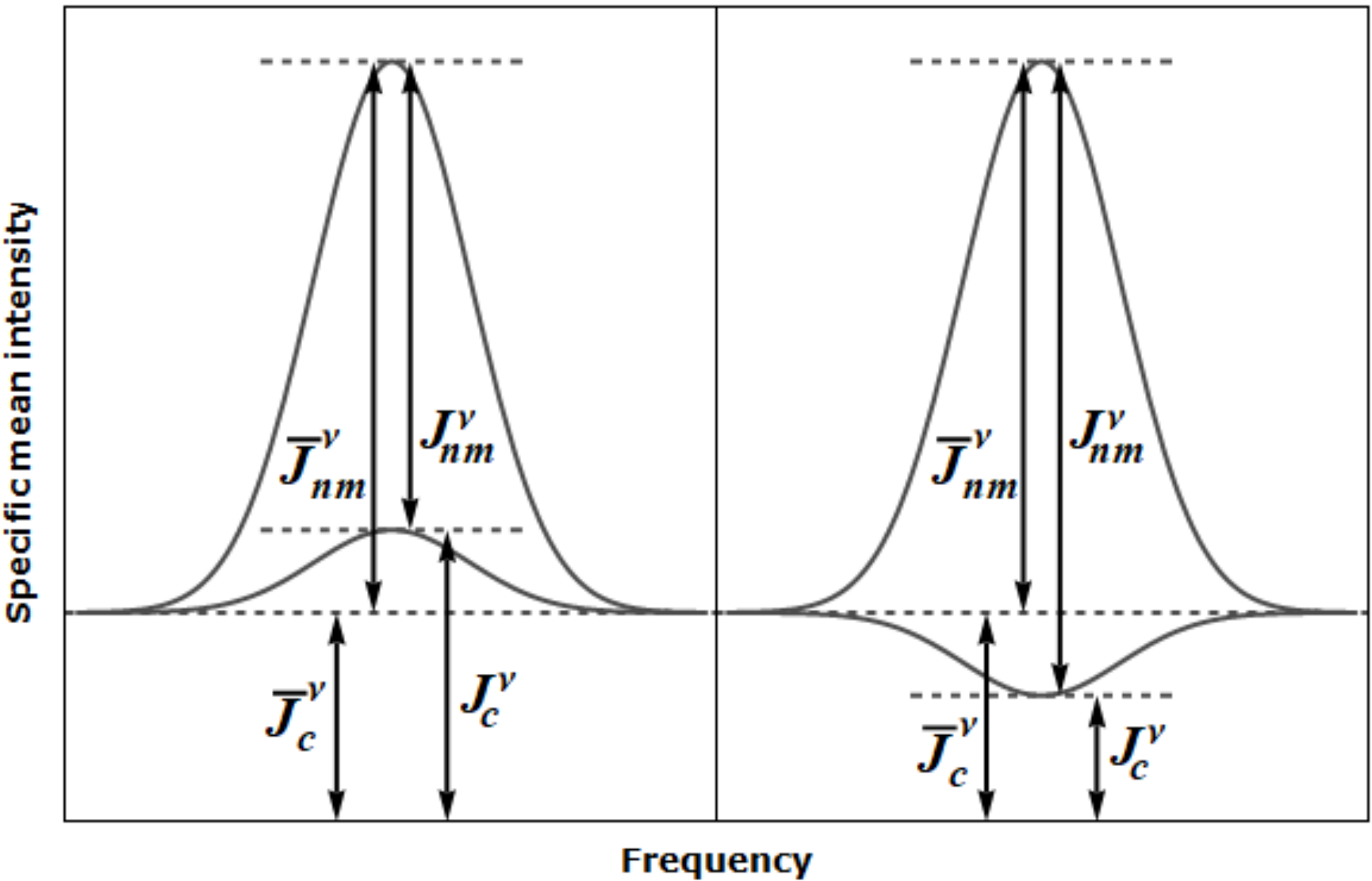}
		\caption{The effects of optical depths on the line and continuum radiation within a spectral line. The left panel shows the situation for $\tau_\nu<-1$ and the right panel shows the situation for $\tau_\nu>1$}
\label{fig:line_effects}
\end{figure}

If maser action is present in the line then $\bar{J}^\nu_{n+1,n}>J^\nu_{n+1,n}$. This occurs because the continuum photons will also be enhanced by stimulated emission and contribute to the observable intensity at the line centre. This scenario is illustrated in the left panel of Fig. \ref{fig:line_effects}. In a non-masing line with a large positive optical depth ($\tau_{\nu} > 1$), $\bar{J}^\nu_{n+1,n}<J^\nu_{n+1,n}$ so that the observable line intensity will underestimate the contribution of the line photons to the total emission observed in the line, as shown in the right panel of Fig. \ref{fig:line_effects}. This discrepancy increases as the magnitude of the optical depth increases, or, for the case illustrated in Figs. \ref{fig:bump_dist} and \ref{fig:bump_dist_obs}, as the path length is increased.

\begin{figure}
	\includegraphics[width=\columnwidth]{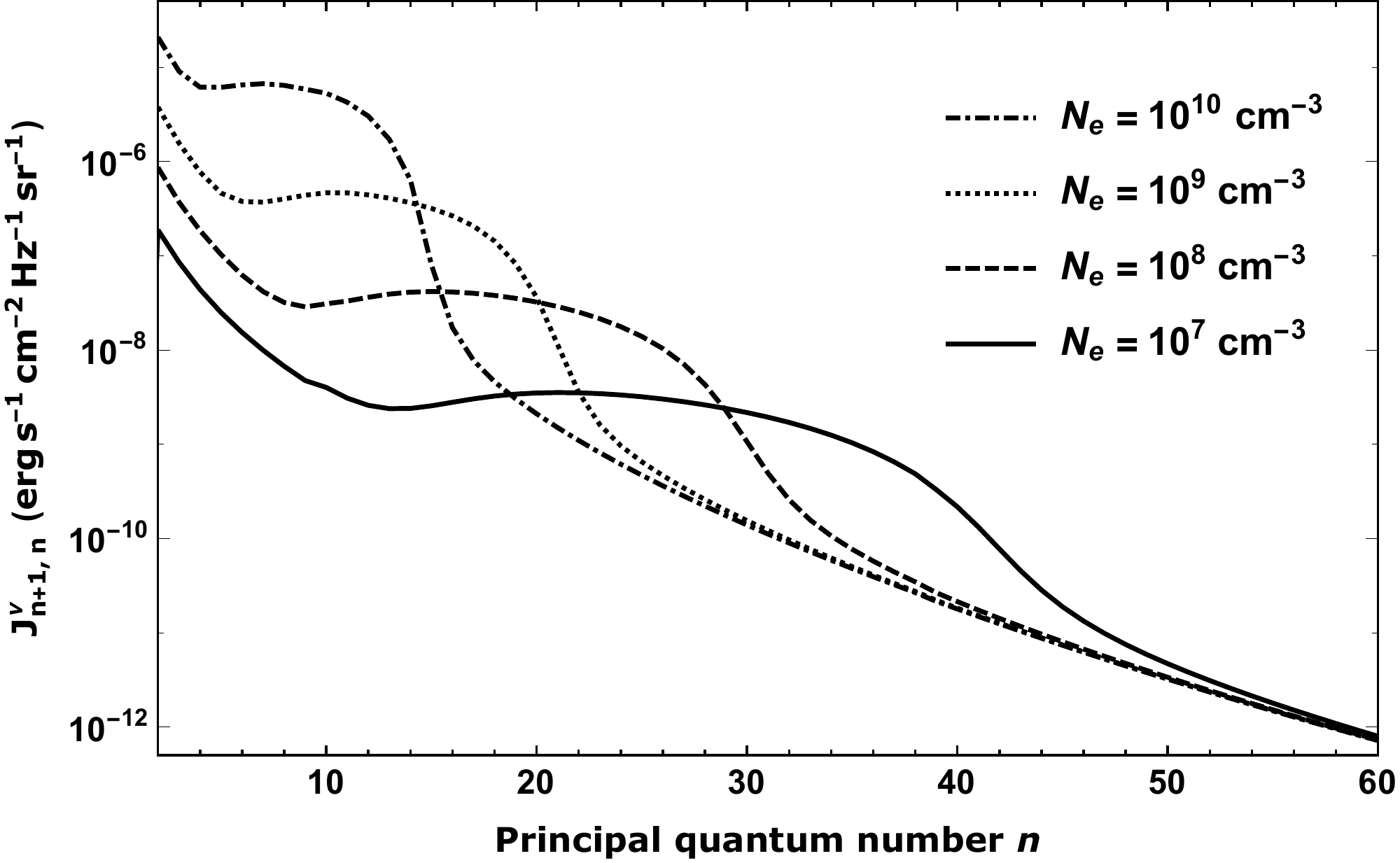}
		\caption{Emitted spectrum of H$n\alpha$ transitions for
masing region at electron temperature $T_\mathrm{e}=10^4$\,K
for a range of electron densities.}
\label{fig:bump_ne}
\end{figure}

Fig. \ref{fig:bump_ne} illustrates the effect of the electron density
on the emitted spectrum of H$n\alpha$ lines. The kinetic temperature is the same for
all of the models shown ($T_\mathrm{e}=10^4$\,K) and
the path length is chosen to show a pronounced bump for each case.
The H$n\alpha$ lines that exhibit masing are mostly determined by
the electron density, with masing occurring at lower $n$-levels for
higher densities. The behaviour of the the H$n\alpha$ lines in the
optically thick regime (high $n$) is independent of density.

The effects of temperature on the emitted spectrum of a hydrogen maser
region is much less pronounced than that of electron density. The values of $n$ where maser action occurs increases slightly towards higher $n$-levels as the temperature increases.

\subsection{Comparison with optically thin and $n$-model results}
\label{subsec:optinvs}

Departure coefficients calculated assuming optically thin conditions (C$^3$ models) are often used as a first approximation when doing HRL maser calculations, see for example \citet{1996Strelnitski} and \citet{2013Baez-Rubio}. It has also been argued that due to the high densities at which HRL masers occur, it is reasonable to neglect the $l$-structure of the atoms and use the results of an $n$-model \citep{1996Strelnitski,2000Hengel}.

\begin{figure}
	\includegraphics[width=\columnwidth]{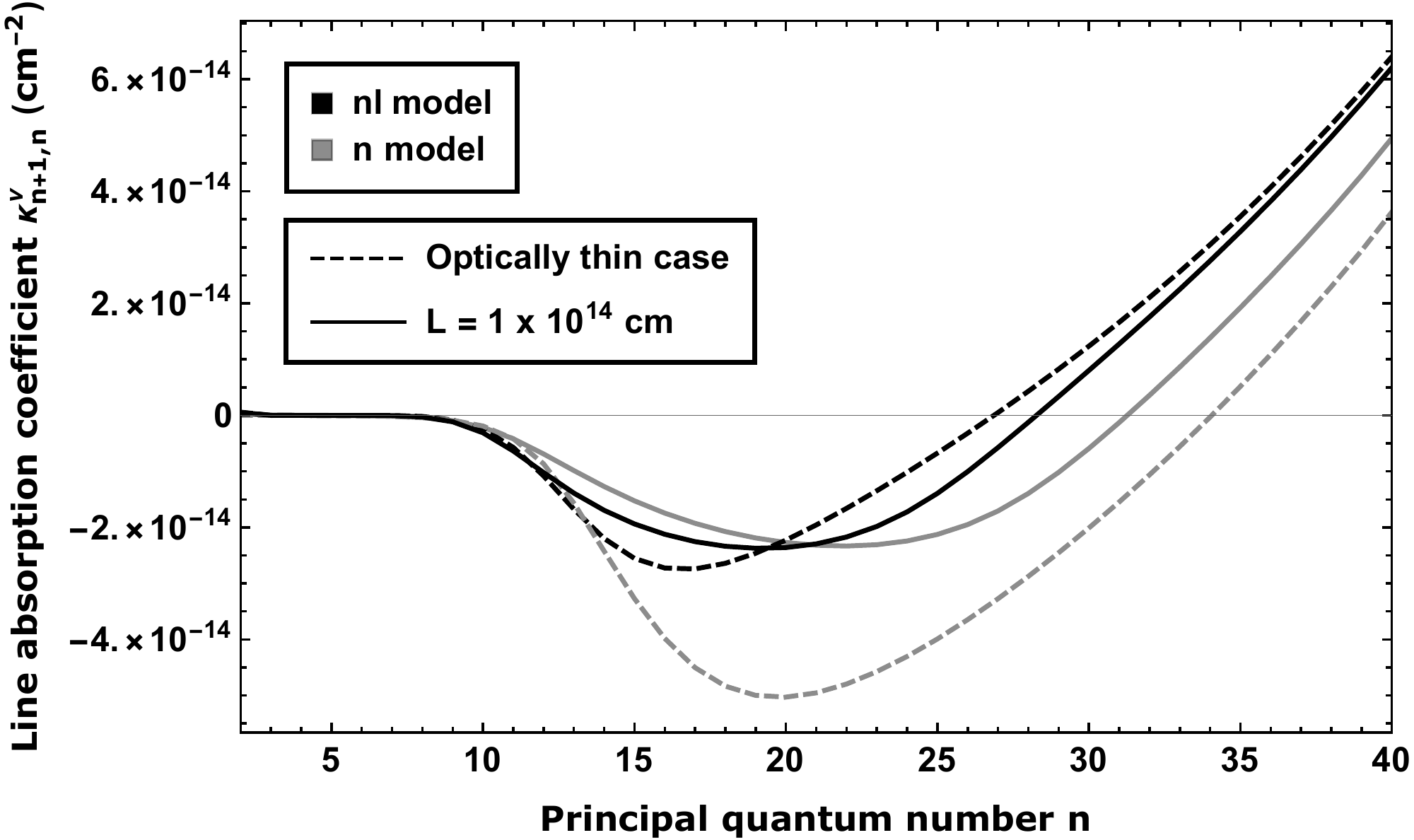}
		\caption{Absorption coefficients at line centre for H$n\alpha$ transition lines as calculated by different model assumptions as indicated in the legend. The electron temperature was set to $T_\mathrm{e}=10^4$\,K and the electron density to  $N_\mathrm{e}=10^8$\cmc.}
\label{fig:kappa_comp}
\end{figure}

Fig. \ref{fig:kappa_comp} shows the line absorption coefficients at line centre of H$n\alpha$ as calculated using the level populations from models with different assumptions. Using the optically thin $n$-model results overestimates the maser gain by a factor of a few. Both the $n$-models (C$^3$ and EPA) will overestimate the range of  H$n\alpha$ transitions that can exhibit masing towards higher $n$-levels for a given set of conditions. The two $nl$-model results are similar, and deciding which to use will depend on the accuracy required for a specific application. The EPA $nl$-model predictions are more accurate than the C$^3$ $nl$-model results in two ways, namely the $n$-level where the most maser gain will be seen and the range of lines over which maser action can occur.
\begin{figure}
	\includegraphics[width=\columnwidth]{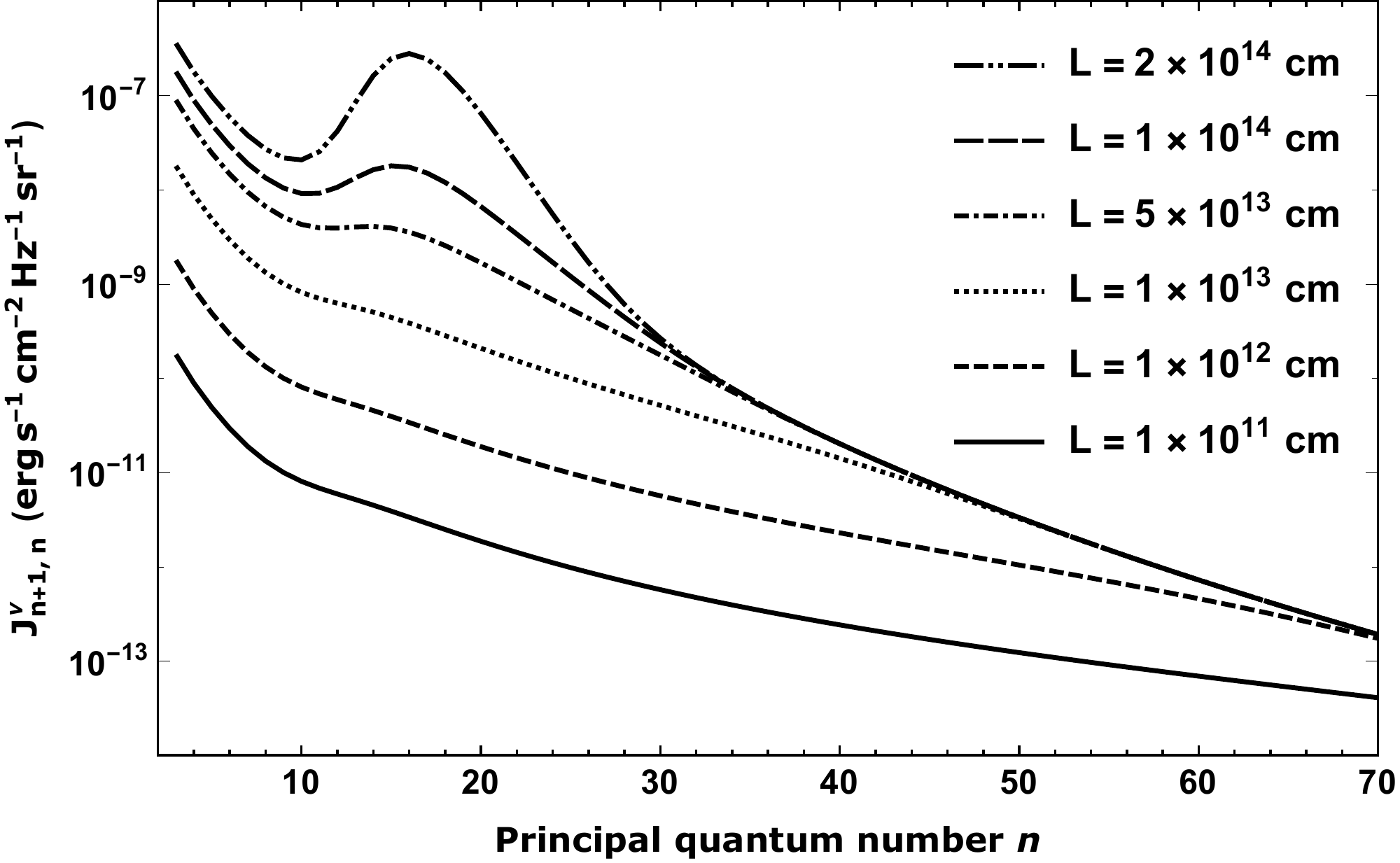}
		\caption{Similar to Fig. \ref{fig:bump_dist}, but using $b_{nl}$ values calculated in the optically thin approximation for all path lengths. }
\label{fig:opthinbnls}
\end{figure}

Fig. \ref{fig:opthinbnls} shows the predicted line intensities if level populations from a C$^3$ $nl$-model are used as opposed to those from the EPA $nl$-model that was illustrated in Fig. \ref{fig:bump_dist}. In the C$^3$ case, the maximum maser gain is shifted to a lower $n$-level and the bump feature is much sharper because the level populations have not been changed by the diffuse radiation. Using the C$^3$ $n$-model results show a similar trend, but with the bump feature much exaggerated, as expected from the large magnitude of the line absorption coefficient shown in Fig. \ref{fig:kappa_comp}.

\citet{1996Strelnitski} discuss the possibility of saturation in
higher frequency lines ``attracting'' lower frequency lines to exhibit
masing. Saturation in one of the masing lines, say $2\rightarrow 1$,
will cause a decrease in the upper level of the transition, level
$2$. This will increase the inversion between levels $3$ and $2$
and therefore the $3\rightarrow2$ line will increase in intensity.
If this process is effective enough, masing will start in the $3\rightarrow2$
transition and the mechanism can diffuse out to even higher levels.

\begin{figure}
	\includegraphics[width=\columnwidth]{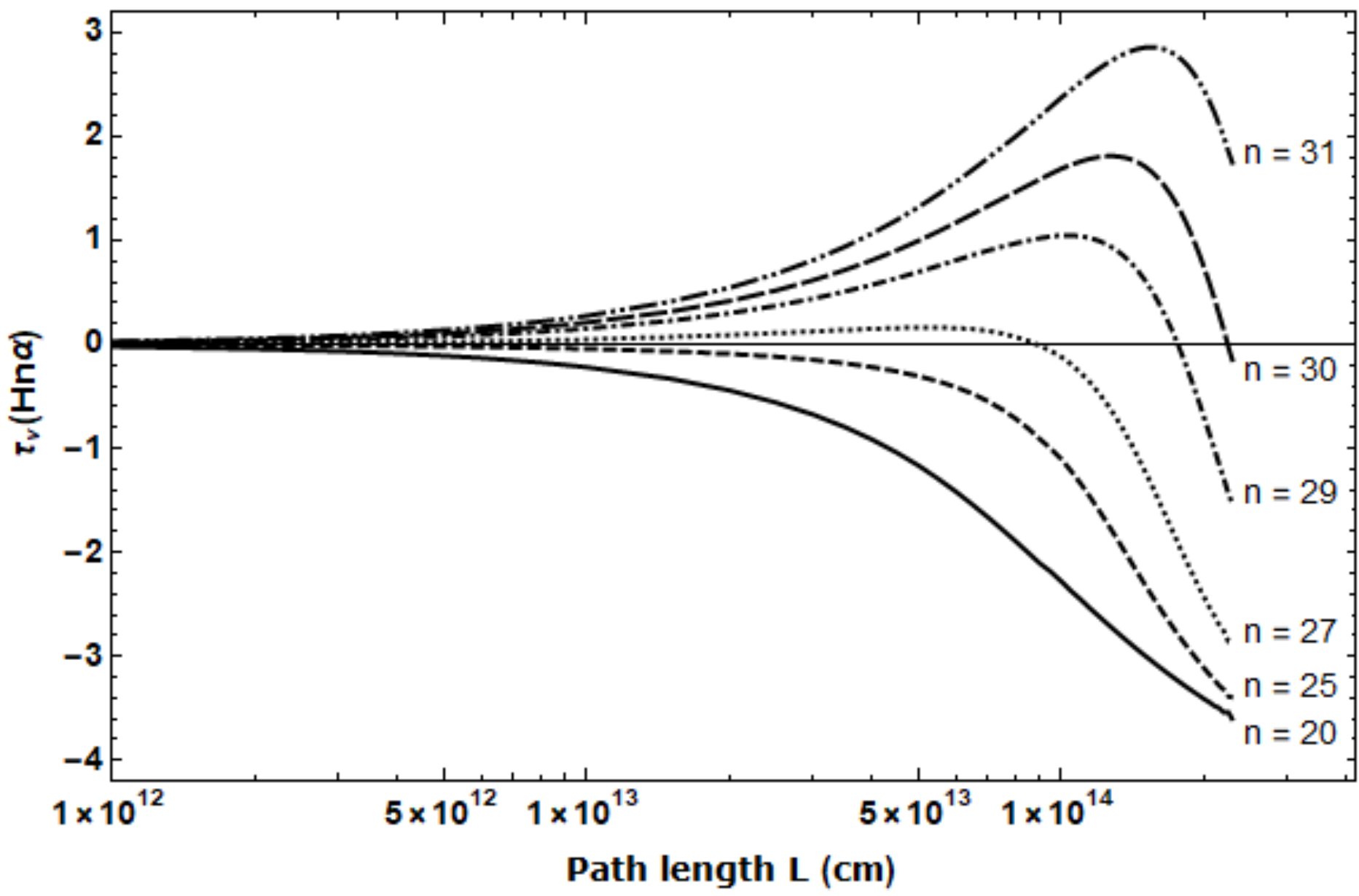}
		\caption{The change in optical depths with increasing
path length for selected H$n\alpha$ transitions under conditions
of $T_\mathrm{e}=10^4$\,K and $N_\mathrm{e}=10^8$\cmc.}
\label{fig:tau_vs_L}
\end{figure}

This behaviour is seen in the EPA model and is illustrated in Fig. \ref{fig:tau_vs_L}, which shows
how the total optical depths at certain H$n\alpha$ line frequencies changes as $L$ is increased.
The optical depths at the H$27\alpha$ to H$30\alpha$ transition frequencies
are positive for small values of $L$ and become negative and
start to exhibit masing behaviour at some larger value of $L$. This
occurs when the H$n\alpha$ transitions for lower $n$-levels' degree
of saturation have increased enough to increase the inversion of higher
transitions, expanding the range of lines where maser action is possible.

\subsection{Comparison with observations}
\label{subsec:comp-obs}

We compare results from the model with observations obtained by \citet{1995Thum} and \citet{1998Thum}
for MWC 349A\@. This is the best studied hydrogen maser region at present
and intensity data is available over a large range of frequencies.
\citet{2018Aleman} more recently published observations of the nebula
Mz 3 where hydrogen masing is also observed. The line intensity ratios
of \citet{2018Aleman} are very similar to those of \citet{1998Thum},
but the observations cover a much smaller frequency range than for MWC 349A\@. The limited observations
do not show the complete bump structure in the H$n\alpha$ spectrum,
so the model presented here cannot be used to constrain the physical
parameters adequately.

The three sets of data that were used to find a reasonable model are the H$n\alpha$ fluxes from \citet{1998Thum}, the $\beta/\alpha$ ratios for which the lines are close in $n$ from the same paper, and the $\beta/\alpha$ ratios for which the lines are close in frequency from \citet{1995Thum}. An H$n\beta$ line represents the transition $n+2\rightarrow n$. The results from the EPA model that gave a reasonable fit considering all three sets of data occur for $T_\mathrm{e} = 11\,000$\,K, 
$N_\mathrm{e} = 7\times10^7$\cmc\ and $L = 2.7\times10^{14}$\,cm. These parameters are similar to the values obtained by \citet{2013Baez-Rubio} for the disc of MWC 349A which is the putative source of the maser emission. It is possible to get a model that matches one of the sets of data better than the chosen model, but it would be very inaccurate for the others. The aim of this fit is not to determine exact properties of MWC 349A, but to show that the EPA model gives results consistent with current observations.

\begin{figure}
	\includegraphics[width=\columnwidth]{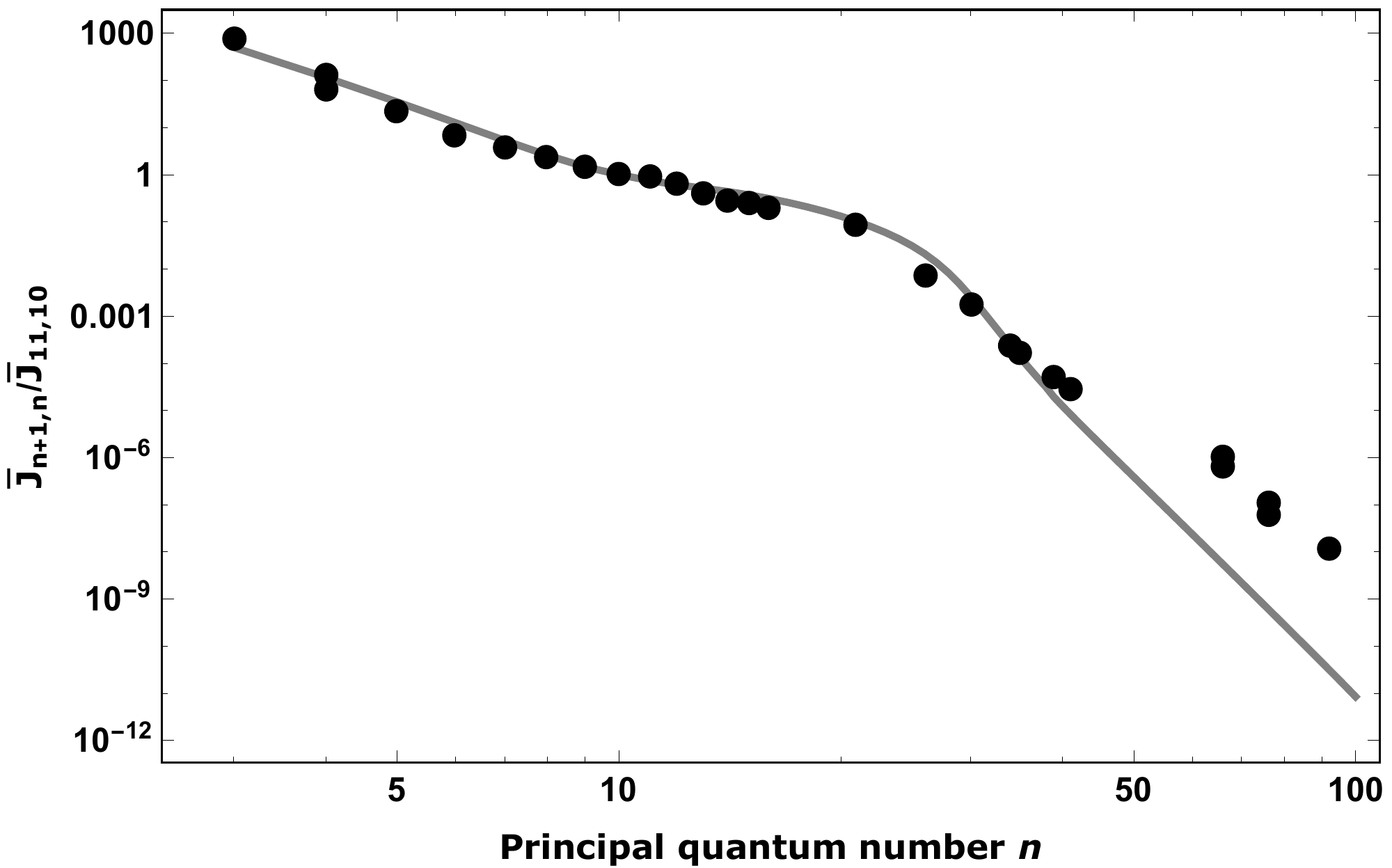}
		\caption{Results from the model (solid line) compared to data from \citet{1998Thum} (circles)
for integrated line intensities of H$n\alpha$ transitions relative to H$10\alpha$. $T_\mathrm{e} = 11\,000$\,K,
$N_\mathrm{e} = 7\times10^7$\cmc\ and $L = 2.7\times10^{14}$\,cm.}
\label{fig:Thum fit}
\end{figure}

Fig. \ref{fig:Thum fit} compares the data of \citet{1998Thum}
and the chosen fit produced by the current model. The figure shows the observable
integrated line intensities $\bar{J}_{n+1,n}$ obtained from equation~(\ref{eq:opthin_Inm}) of H$n\alpha$ transitions relative to the H$10\alpha$ intensity. The model greatly underestimates the observable intensities for lines with $n>40$. It is believed that observed $\alpha$-lines with $n>38$ originate in the outflow and not in the disc where the masers are formed \citep{1992Planesas}. The conditions in the outflow are different to those in the disc.

\begin{figure}
	\includegraphics[width=\columnwidth]{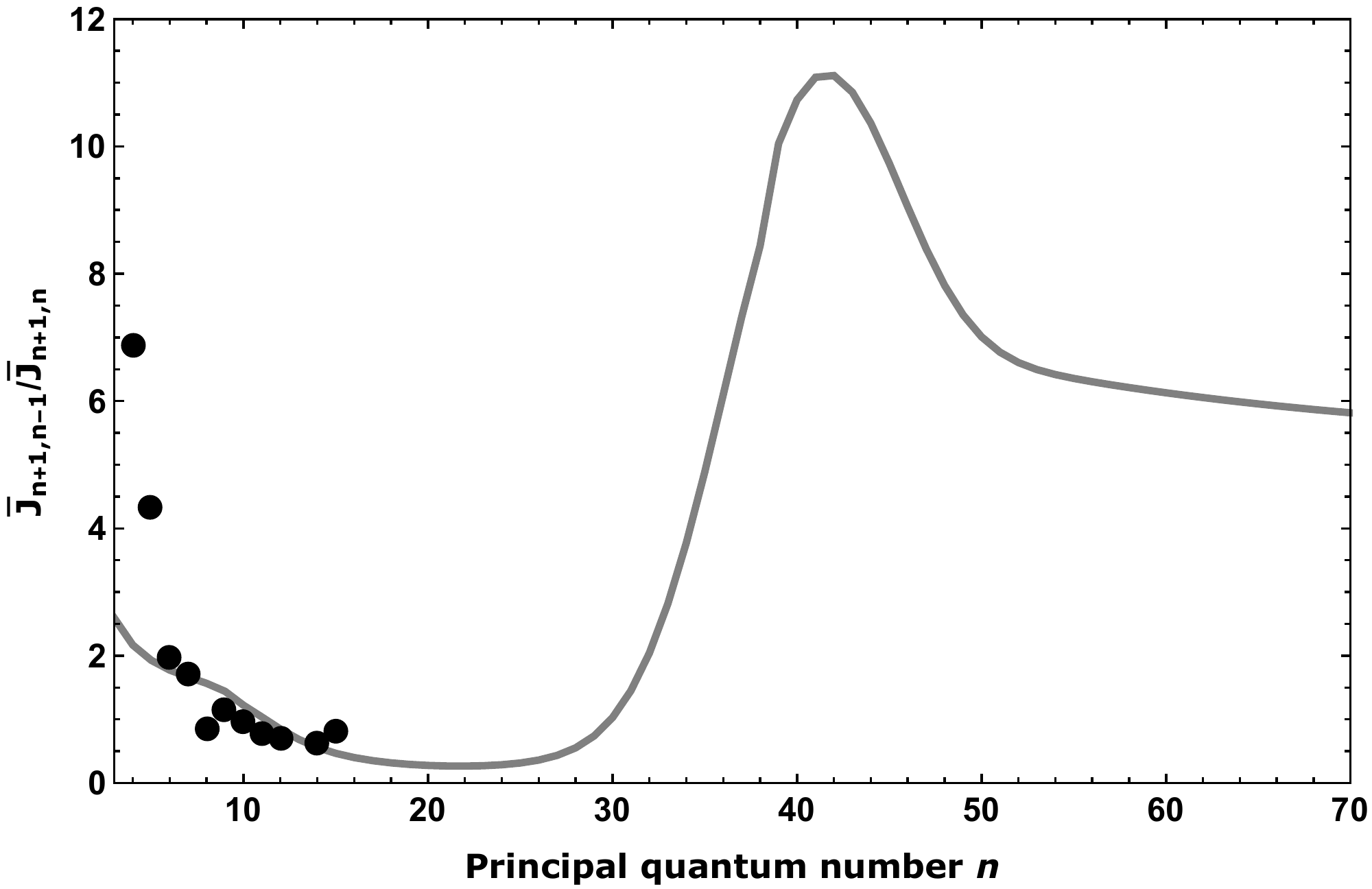}
		\caption{The ratio of H$n\alpha$ and H$n\beta$
lines with the same upper level. Circles show observations from \citet{1998Thum},
the solid line shows results from the same model as in Fig. \ref{fig:Thum fit}.}
\label{fig:Thum_beta_alpha}
\end{figure}

Fig. \ref{fig:Thum_beta_alpha} shows the ratios of the integrated observed H$n\beta$
and H$n\alpha$ lines with the same upper level (black circles) for the model (solid line)
shown in Fig. \ref{fig:Thum fit}. Agreement between the observations
and model is good, except for the lowest $n$ levels. Because the population
of the ground level is set artificially to a fixed degree of ionization,
the results for the lowest $n$ levels are not reliably determined by the model. The figure also shows the model predictions for the ratio for $n>16$. The results for $n>40$ will probably not be observed, since the measured lines from MWC 349A at those frequencies are not emitted from the disc. A list of H$n\alpha$, H$n\beta$ and H$n\gamma$ lines that occur within the frequency bands of ALMA is given in Table~\ref{tab:alma}.

A comparison between our model results for $T_\mathrm{e} = 11\,000$\,K,
$N_\mathrm{e} = 7\times10^7$\cmc\ and $L = 2.7\times10^{14}$\,cm. and the observations of \citet{1995Thum} for $\beta/\alpha$ ratios that are close in frequency is presented in Table \ref{tab:thumratios}. Most ratios could be fitted within the given ranges of \citet{1995Thum}. The worst outlier is the ratio 48$\beta$/40$\alpha$, which is consistent with the model's inability to fit lines with $n>40$.

\begin{table}
	\centering
	\caption{Model results compared to observed ratios of H$n\beta$/H$n\alpha$ pairs that are close in frequency as published by \citet{1995Thum}.}
	\label{tab:example_table}
	\begin{tabular}{lcc} 
		\hline
		Lines & Model ratio (\%) & Observed ratio (\%)\\
		\hline
		37$\beta$/30$\alpha$ & 9.9  & $< 12$\\
        33$\beta$/26$\alpha$ & 4.2  & $< 10$\\
    	39$\beta$/31$\alpha$ & 9.3  & $12 \pm 3$\\
        33$\beta$/26$\alpha$ & 4.2  & $8.5 \pm 1.5$\\
        32$\beta$/26$\alpha$ & 5.6  & $4.5 \pm 1.5$\\
        45$\beta$/36$\alpha$ & 15  & $< 18$\\
        38$\beta$/30$\alpha$ & 7.1  & $6.6 \pm 1.5$\\
        48$\beta$/40$\alpha$ & 31  & $14 \pm 2$\\
		\hline
	\end{tabular}
	\label{tab:thumratios}
\end{table}

\subsection{Ratios of $\alpha$- and $\beta$-lines}

Comparing the intensities of H$n\beta$ lines to those of H$n\alpha$ transitions can provide additional information regarding the physical conditions in the emitting region. \citet{1996StrelnitskiB} concluded that it is preferable to consider $\beta/\alpha$ pairs that are close in $n$-value for masing regions, as opposed to pairs that are close in frequency. This section reviews some general theoretical trends that should be observable in masing regions.

\begin{figure}
	\includegraphics[width=\columnwidth]{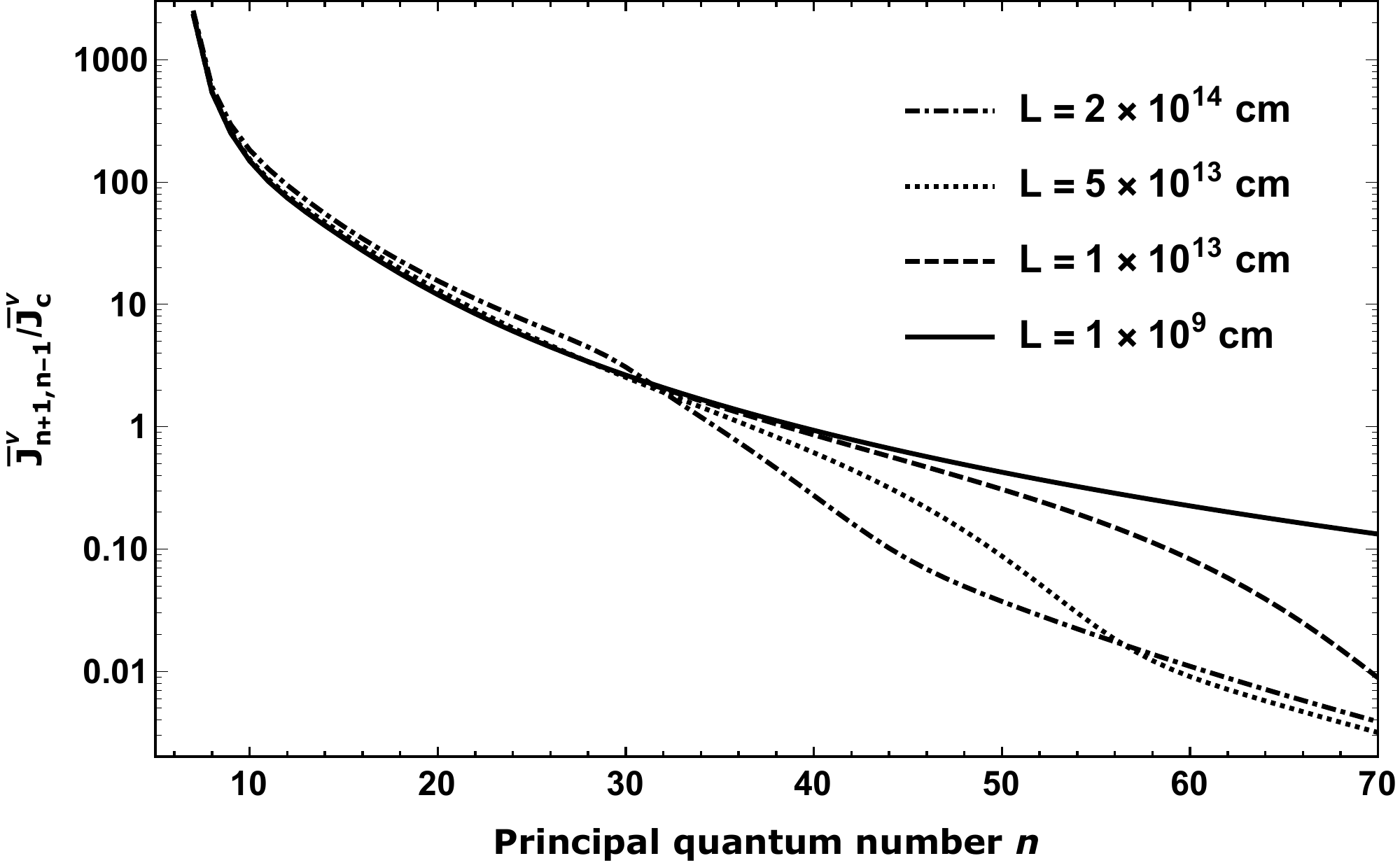}
		\caption{The ratio of the observable line centre $\bar{J}^{\nu}_{n+1,n-1}$ to continuum intensities of H$(n-1)\beta$ lines for a gas with $T_\mathrm{e}=10^4$\,K
and $N_\mathrm{e}=10^8$\cmc\ for various path lengths.}
\label{fig:obs_beta}
\end{figure}

Fig. \ref{fig:obs_beta} is a plot of the ratio of the line intensity $\bar{J}^{\nu}_{n+1,n-1}$ and continuum vs $n$ which gives an indication of the observability of H$(n-1)\beta$ transitions (i.e.\ $n+1 \rightarrow n-1$) for a masing gas. The solid line in the graph is an approximation of the optically thin case. For larger optical depths, the $\beta$-lines are slightly enhanced in the same $n$-region as the $\alpha$-lines, but they are not technically masing. As the path length is increased for the higher $n$-transitions, the observable line to continuum ratio decreases.

\begin{figure}
	\includegraphics[width=\columnwidth]{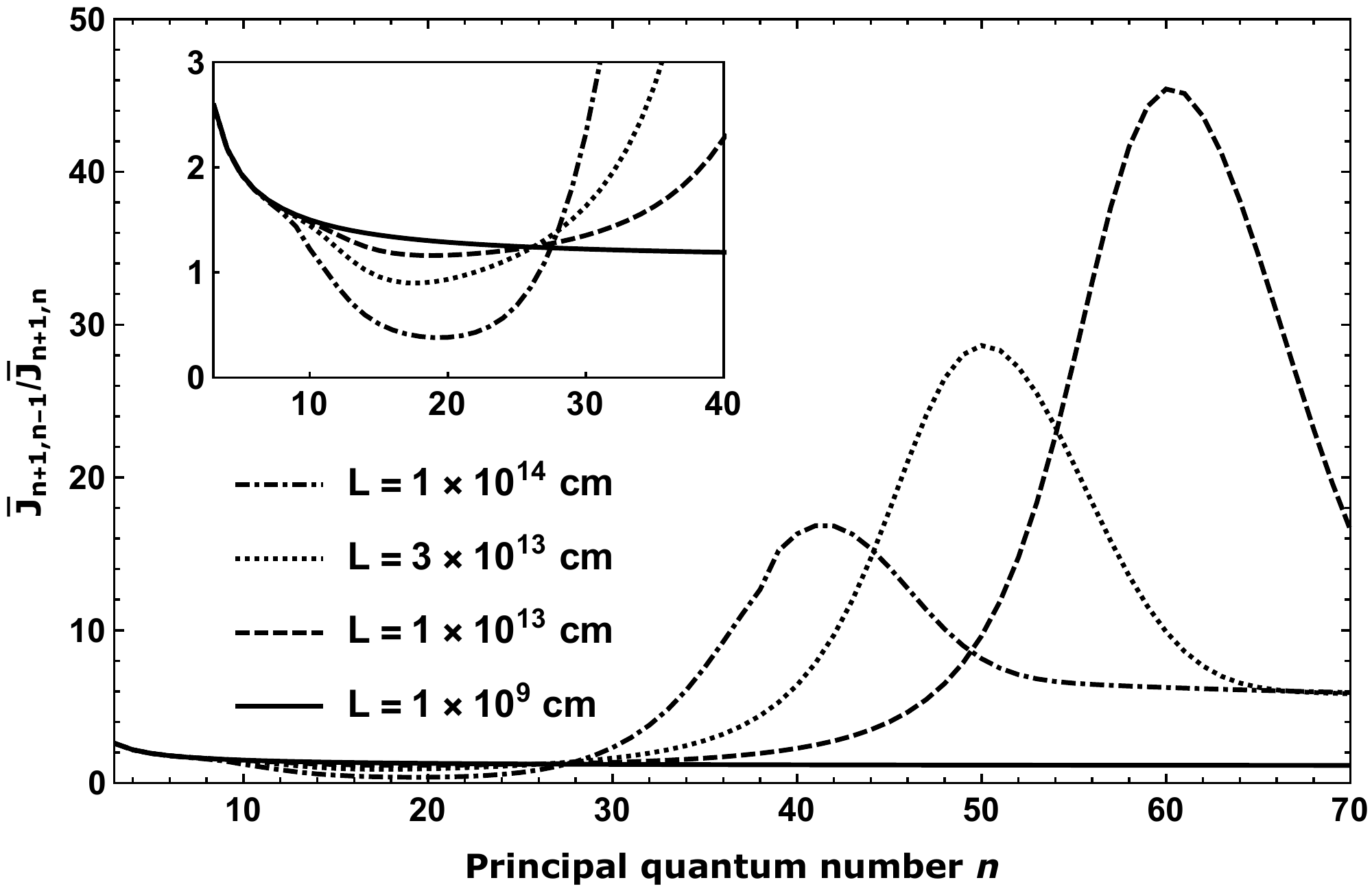}
		\caption{The model ratios of $\beta$- and $\alpha$-lines
with the same upper level for a masing gas with $T_\mathrm{e}=10^4$\,K
and $N_\mathrm{e}=10^8$\cmc\ for various path lengths. The inset magnifies the behaviour at small $n$. }
\label{fig:gen_alpha_beta_ratio}
\end{figure}

For a single emitting atom, the intensity of the $\alpha$-lines will
always be greater than or equal to the intensity of the $\beta$-lines,
because the $\alpha$-transitions have a larger transition probability than the $\beta$-transitions.
But when considering a cloud, the difference in optical depths of
the two transitions plays a role. The optical depths of the $\alpha$-transitions
are much larger than those of the $\beta$-transitions for the same
upper level, so that the $\alpha$-lines will always have a higher
optical depth than their $\beta$ counterparts in the region where
masing is not present.

Fig. \ref{fig:gen_alpha_beta_ratio} shows the behaviour of the observable integrated intensities $\bar{J}_{n+1,n-1}/\bar{J}_{n+1,n}$ as a function of $n$ for the case $T_\mathrm{e}=10^4$\,K and $N_\mathrm{e}=10^8$\cmc\ for different values of $L$ as produced by the EPA model. In the region 
where the H$n\alpha$ lines exhibit masing (inset in Fig.\ \ref{fig:gen_alpha_beta_ratio}), the 
ratio is decreased as $L$ increases and the $\alpha$-lines outshine the
non-masing $\beta$-lines. At high values of $n$ the situation is reversed with the $\beta$-lines 
becoming more intense than the $\alpha$-lines.

For example, for the $L=10^{14}$\,cm case shown in Fig.
\ref{fig:gen_alpha_beta_ratio} the line optical depths at line centre are $\tau_{41,40}=6$
and $\tau_{41,39}=0.8$. This means that the characteristic path length
of a $\beta$-photon is larger than that of an $\alpha$-photon by
a factor of about $7.5$ and we are observing $\beta$-photons coming
from 7.5 times deeper in the cloud than the $\alpha$-photons. Therefore,
the observed intensities of the $\beta$-lines are much brighter than
those of the $\alpha$-lines if one or both of the transitions become
optically thick. If both of the transitions are optically thick, the 
integrated intensity ratio of the two lines will tend to $\sim 5$, regardless 
of physical conditions. 

The bump feature that emerges in Fig. \ref{fig:gen_alpha_beta_ratio} between the masing region and where both lines are optically thick is due to the interaction between the line and continuum opacities for the observable intensities. The maximum of the bump occurs at a point where $\tau_\mathrm{c} \approx 1.5$ for the $\beta$ lines.


\section{Conclusions}
\label{sec:Conclusions}

Hydrogen recombination masers are a relatively new field of study
with only a handful of examples detected so far. There are some important
differences between molecular and atomic masers, both on the macro
scale like the environments where they form, and at the atomic level
like the pumping mechanism and interaction of many masing lines. The
theoretical framework for these objects is still developing and the
aim of this paper is to contribute to our understanding by constructing
a theoretical model that specifically focuses on the atomic process
rather than the geometry and kinematics.

The modeling of masers has some inherent complexities, since both
the local level populations of the masing species and the non-local
radiative transfer of the line photons have to be solved simultaneously
in principle. Simplifying assumptions are often employed, for example
the EPA used here. The EPA has limitations, but it is also a very
useful tool to gain insight into overall emission from a cloud.

The effects of incorporating angular momentum changing collisions
into an EPA model for hydrogen was shown. At the high electron
densities where hydrogen masers form, the free-free emission from
electrons will also effect the level populations of hydrogen.

A model for hydrogen recombination masers using the EPA have been
constructed to evaluate the general behaviour of hydrogen emission
from clouds with conditions where masing is possible. The observable
effects of line of sight path length, electron temperature and electron
density on the intensities of H$n\alpha$ emission
lines have been investigated and discussed. The model results for
varying path length corresponds well to our current understanding
of how masers grow with increasing path length. The electron density
has the biggest effect on which transitions will exhibit masing, whereas
temperature has a larger effect on the behaviour of low frequency
non-masing lines.

A fit of the model results was done to observations of the region
MWC 349A where masing in hydrogen was first discovered. A good fit
was obtained for a large range of frequencies with physical parameters
in line with what other authors have obtained.

The behaviour of the ratios of H$n\alpha$ and H$n\beta$ lines that
form from the same upper level have been examined. A fit was done
to high frequency observations that are available and model predictions
for lower frequency transitions are shown.




\bibliographystyle{mnras}
\bibliography{mnras_masers} 



\appendix

\section{Recombination lines observable with ALMA}

\begin{table*}
	\centering
	\caption{List of H$n\alpha$, H$n\beta$ and H$n\gamma$ lines that occur in the ALMA frequency bands.}
	\label{tab:alma}
	\begin{tabular}{clcc|clcc} 
		\hline
		 Upper & Line & $\nu$  & ALMA & Upper&Line &$\nu$  & ALMA\\
		 level &      & ($10^{11}$ Hz) & band  &level &     & ($10^{11}$ Hz) & band\\
		\hline

20	& H19$\alpha$	& $ 8.880$&	10	&38	& H35$\gamma$	& $ 4.071$&	8  \\
22	& H21$\alpha$	& $ 6.624$&	9	&38	& H36$\beta$	& $ 2.600$&	6  \\
25	& H24$\alpha$	& $ 4.475$&	8	&39	& H37$\beta$	& $ 2.400$&	6  \\
26	& H24$\beta$	& $ 8.444$&	10	&40	& H37$\gamma$	& $ 3.468$&	7  \\
26	& H25$\alpha$	& $ 3.969$&	8	&40	& H38$\beta$	& $ 2.220$&	6  \\
27	& H26$\alpha$	& $ 3.536$&	7	&41	& H38$\gamma$	& $ 3.210$&	7  \\
28	& H26$\beta$	& $ 6.700$&	9	&41	& H39$\beta$	& $ 2.058$&	5  \\
28	& H27$\alpha$	& $ 3.164$&	7	&42	& H39$\gamma$	& $ 2.978$&	7  \\
29	& H28$\alpha$	& $ 2.843$&	7	&42	& H40$\beta$	& $ 1.911$&	5  \\
30	& H27$\gamma$	& $ 8.570$&	10	&43	& H40$\gamma$	& $ 2.767$&	7  \\
30	& H29$\alpha$	& $ 2.563$&	6	&43	& H41$\beta$	& $ 1.777$&	5  \\
31	& H29$\beta$	& $ 4.882$&	8	&44	& H41$\gamma$	& $ 2.576$&	6  \\
31	& H30$\alpha$	& $ 2.319$&	6	&44	& H42$\beta$	& $ 1.656$&	5  \\
32	& H29$\gamma$	& $ 6.987$&	9	&45	& H42$\gamma$	& $ 2.402$&	6  \\
32	& H30$\beta$	& $ 4.424$&	8	&45	& H43$\beta$	& $ 1.546$&	4  \\
32	& H31$\alpha$	& $ 2.105$&	5	&46	& H43$\gamma$	& $ 2.244$&	6  \\
33	& H30$\gamma$	& $ 6.341$&	9	&46	& H44$\beta$	& $ 1.445$&	4  \\
33	& H31$\beta$	& $ 4.022$&	8	&47	& H44$\gamma$	& $ 2.099$&	5  \\
33	& H32$\alpha$	& $ 1.917$&	5	&47	& H45$\beta$	& $ 1.352$&	4  \\
34	& H32$\beta$	& $ 3.667$&	7	&48	& H45$\gamma$	& $ 1.966$&	5  \\
34	& H33$\alpha$	& $ 1.750$&	5	&48	& H46$\beta$	& $ 1.268$&	4  \\
35	& H33$\beta$	& $ 3.352$&	7	&49	& H46$\gamma$	& $ 1.844$&	5  \\
35	& H34$\alpha$	& $ 1.602$&	4	&50	& H47$\gamma$	& $ 1.733$&	5  \\
35	& H34$\alpha$	& $ 1.602$&	5	&51	& H48$\gamma$	& $ 1.630$&	4  \\
36	& H33$\gamma$	& $ 4.823$&	8	&51	& H48$\gamma$	& $ 1.630$&	5  \\
36	& H34$\beta$	& $ 3.073$&	7	&52	& H49$\gamma$	& $ 1.535$&	4  \\
36	& H35$\alpha$	& $ 1.470$&	4	&53	& H50$\gamma$	& $ 1.447$&	4  \\
37	& H34$\gamma$	& $ 4.425$&	8	&54	& H51$\gamma$	& $ 1.366$&	4  \\
37	& H35$\beta$	& $ 2.823$&	7	&55	& H52$\gamma$	& $ 1.290$&	4  \\
37	& H36$\alpha$	& $ 1.353$&	4	&	& 	& &	 \\			
		\hline
	\end{tabular}
\end{table*}


\bsp	
\label{lastpage}
\end{document}